\documentclass[aps,pra,twocolumn]{revtex4-1} %nofootinbib
\usepackage{verbatim}
\usepackage{dsfont} 
\usepackage{amsmath}
\usepackage{graphicx}
\usepackage{mathtools} 
\usepackage{bm} 
\usepackage{amssymb} 
\usepackage{enumerate} 
\usepackage{hyperref} 
\hypersetup{
    colorlinks=true,
    linkcolor=blue,
    citecolor=blue,
    unicode=true,          
}

\def\equationautorefname~#1\null{#1\null}

\newcommand{\ket}[1]{\left| #1 \right\rangle}
\newcommand{\bra}[1]{\left\langle #1 \right|}
\newcommand{\braket}[1]{\left\langle #1 \right\rangle}
\newcommand{\ketbra}[2]{\left|#1 \rangle \langle #2 \right|}

\begin{document}

%\preprint{}

\title{Weight of informativeness, state exclusion games and excludible information}


\author{Andr\'es F. Ducuara$^{1,2}$, Paul Skrzypczyk$^{3}$} %$^{1,2}$
%\thanks{}
%\affiliation{}

\address{$^{1}$Quantum Engineering Centre for Doctoral Training, \\H. H. Wills Physics Laboratory and Department of Electrical \& Electronic Engineering, University of Bristol, BS8 1FD, UK}

\address{$^{2}$Quantum Engineering Technology Labs, H. H. Wills Physics Laboratory and \\
Department of Electrical \& Electronic Engineering, University of Bristol, BS8 1FD, UK.}

\address{$^{3}$H.H. Wills Physics Laboratory, University of Bristol, Tyndall Avenue, Bristol, BS8 1TL, United Kingdom}


\date{\today}

\begin{abstract}
We consider the quantum resource theory of measurement informativeness and introduce a weight-based quantifier of informativeness. We show that this quantifier has operational significance from the perspective of quantum state exclusion, by showing that it precisely captures the advantage a measurement provides in minimising the error in this game. We furthermore  introduce information theoretic quantities related to exclusion, in particular the notion of excludible information of a quantum channel, and show that for the case of quantum-to-classical channels it is determined precisely by the weight of informativeness. This establishes a three-way correspondence which sits in parallel to the recently discovered correspondence in quantum resource theories between robustness-based quantifiers, discrimination games, and accessible information. We conjecture that the new correspondence between a weight-based quantifier and an exclusion-based task found in this work is a generic correspondence that holds in the context of quantum resource theories. 
\end{abstract} 

%\keywords{}

\maketitle

%%%%%%%%%%
%%%%%%%%%%
%%%%%%%%%%
%%%%%%%%%%
%%%%%%%%%%
%%%%%%%%%%
\section{Introduction}
%%%%%%%%%%
%%%%%%%%%%
%%%%%%%%%%
%%%%%%%%%%
%%%%%%%%%%
%%%%%%%%%%

%\emph{Introduction.---}
The 21st-century is currently witnessing a second quantum revolution which, broadly speaking, aims at harnessing different quantum phenomena for the development of quantum technologies. Quantum phenomena can then be seen as a resource for fuelling quantum information protocols. In this regard, the framework of \emph{Quantum Resource Theories (QRTs)} has been put forward in order to address these phenomena within a common unifying framework \cite{RT_review}. There are several QRTs of different quantum `\emph{objects}' addressing different properties (of the object) as a \emph{resource}. We can then broadly classify QRTs by first specifying the objects of the theory, followed by the property to be harnessed as a resource. In this broad classification there are QRTs addressing quantum objects like: states \cite{QRTE1, QRTE2}, measurements \cite{RT_measurements0, RT_measurements1, RT_measurements2}, correlations \cite{RT_nonlocality, RT_noncontextuality1, RT_noncontextuality2}, steering assemblages \cite{RT_steering}, and channels \cite{RT_channels1, RT_channels2}. Arguably, the most studied QRTs are the ones for states and measurements. On the one hand, QRTs of states address resources such as entanglement \cite{QRTE1, QRTE2}, coherence \cite{RT_coherence0,RT_coherence1}, asymmetry \cite{RT_coherence0}, and athermality \cite{RT_athermality}, among many others \cite{RT_superposition, RoP, RT_nongaussianity, RT_nonmarkovianity1, RT_nonmarkovianity2, RT_RF}. QRTs of measurements on the other hand, address resources such as projective simulability \cite{RT_PS} and informativeness \cite{RoM}.

One of the main goals within the framework of QRTs is to define \emph{resource quantifiers} for abstract QRTs, so that resources of different objects can be quantified and compared in a fair manner. There are different measures for quantifying resources, depending on the type of QRT being considered \cite{RT_review}. In particular, when considering \emph{convex} QRTs, well-studied geometric quantifiers include the so-called \emph{robustness-based} \cite{RoE,GRoE,RoS,RoA,RoC, RoT,RoT2,RT_magic} and \emph{weight-based} \cite{EPR2,WoE,WoS, WoI_MP, RoNL_RoS_RoI,WoAC} quantifiers. Both robustness-based and weight-based resource quantifiers can be defined for general convex QRTs and therefore, all of these resources can be quantified and compared on an equal footing. This has allowed the cross-fertilisation across QRTs, in which results and insights from a particular QRT with an specific resource are being extended to additional resources and families of QRTs \cite{RT_review, RT1, RT2}.

In addition to quantifying the amount of resource present in a quantum object, it is also of interest to develop practical applications in the form of \emph{operational tasks} that explicitly take advantage of specific given resources, as well as to identify adequate resources and quantifiers characterising already existing operational tasks. In this regard, a general correspondence between robustness-based measures and \emph{discrimination-based} operational tasks has recently been established: steering for subchannel discrimination \cite{RoS}, incompatibility for ensemble discrimination \cite{RoI_task, RoI_witnesses, RoI_channels}, coherence for unitary discrimination \cite{RoA} and informativeness for state discrimination \cite{RoM}. This correspondence initially considered for specific QRTs and resources, has been extended to QRT of states, measurements and channels with \emph{arbitrary} resources \cite{RT1, RT2}. Furthermore, it turns out that when considering QRTs of measurements there exists an additional correspondence to single-shot information-theoretic quantities \cite{RoM}. This three-way correspondence, initially considered for the resource of informativeness \cite{RoM}, has been extended to convex QRTs of measurements with \emph{arbitrary} resources \cite{RT2}. 

\begin{figure}[h!]
    \centering
    \includegraphics[scale=0.46]{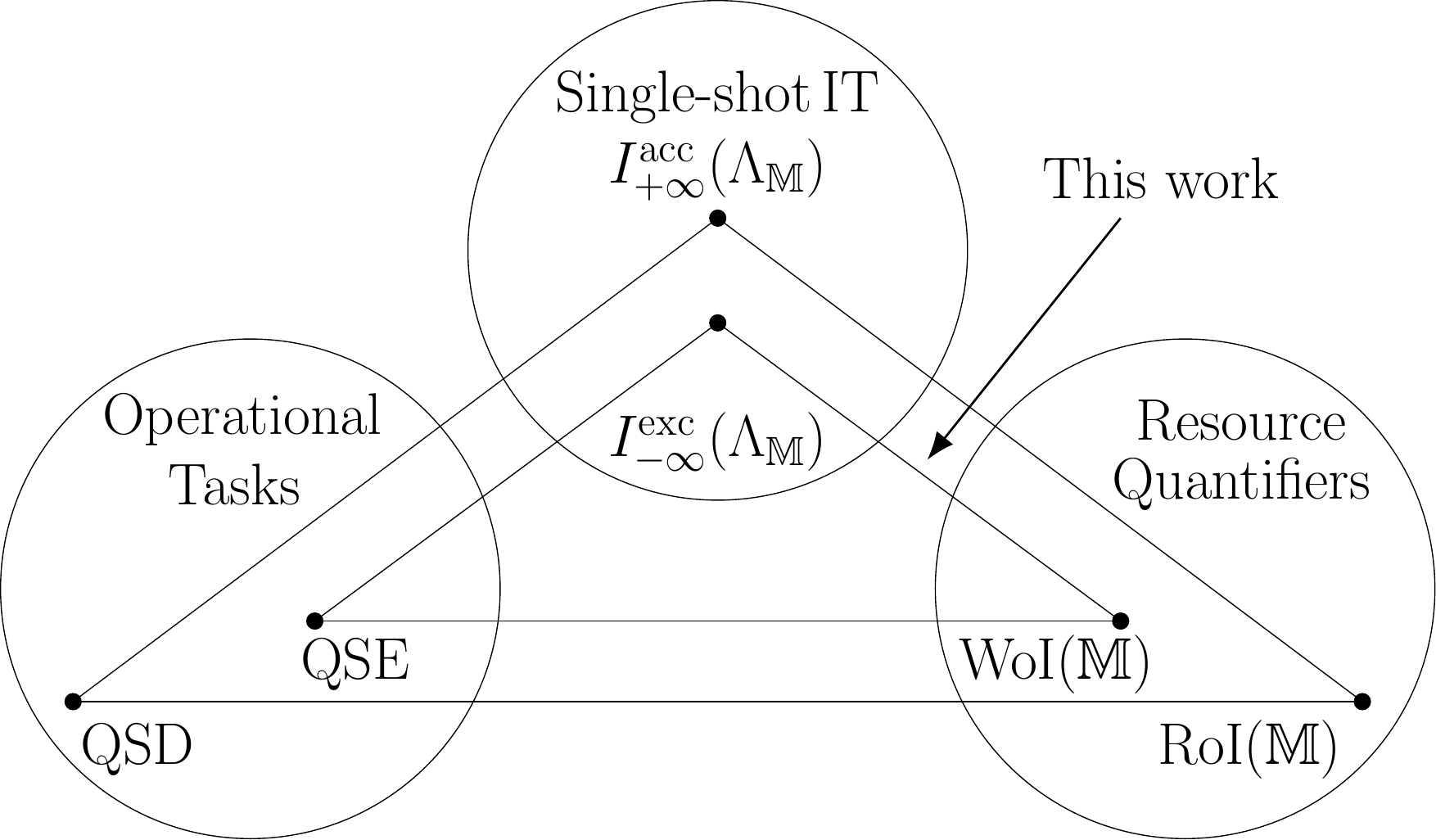}
    \vspace{-0.3cm}
    \caption{Three-way correspondence between: operational tasks, resource quantifiers and single-shot information-theoretic  quantities for the QRT of measurement informativeness. The outer three-way correspondence is linking \cite{RoM}; quantum state discrimination (QSD), robustness of informativeness (RoI) and single-shot accessible 
    information $I_{+\infty}^{\rm acc}(\Lambda_\mathbb{M})$. In this work, we derive a parallel three-way correspondence (inner triangle) linking: weight of informativeness (WoI), quantum state exclusion (QSE) and single-shot excludible information $I_{-\infty}^{\rm exc}(\Lambda_\mathbb{M})$. Definitions of these quantities in the main text.}
    \label{fig:triangle}
\end{figure}

It is then natural to ask whether operational tasks can be devised in which, \emph{weight-based} quantifiers play the relevant role. Surprisingly, in this work we prove that one does not need to design any contrived operational task, but that there are natural operational tasks which are characterised by these weight-based quantifiers, namely, the so-called \emph{exclusion-based} operational tasks. Furthermore, we prove that these weight-based quantifiers for the QRTs of measurements also happen to satisfy a stronger three-way correspondence, establishing again a link to single-shot information-theoretic quantities. In \autoref{fig:triangle} we have a diagrammatic representation of the parallel three-way correspondence found in this work, depicted as the inner triangle. Explicitly, we prove that for convex QRTs of measurements with the resource of \emph{informativeness}, the weight of informativeness quantifies both; the advantage of informative over uninformative measurements in the operational task of \emph{state exclusion} \cite{CES}, and a new type of single-shot accessible information (of the quantum-classical channel induced by a measurement) associated to a novel communication problem.

This parallel three-way correspondence establishes that, in addition to robustness-based quantifiers, weight-based quantifiers also play a relevant role in the characterisation of operational tasks. We conjecture that the \emph{weight-exclusion correspondence} found in this article holds for arbitrary QRTs of different objects beyond those of measurements.  In an upcoming article, we support this conjecture by showing that this is the case for the weight-based resource quantifiers in convex QRTs of states with arbitrary resources and therefore, providing an operational interpretation to these weight-based resource quantifiers.

%%%%%%%%%%%%%%%%%%%%%%%%%%%%%%%
%%%%%%%%%%%%%%%%%%%%%%%%%%%%%%%
%%%%%%%%%%%%%%%%%%%%%%%%%%%%%%%
%%%%%%%%%%%%%%%%%%%%%%%%%%%%%%%
%%%%%%%%%%%%%%%%%%%%%%%%%%%%%%%
%%%%%%%%%%%%%%%%%%%%%%%%%%%%%%%
\section{Convex quantum resource theories and resource quantifiers}
%%%%%%%%%%%%%%%%%%%%%%%%%%%%%%%
%%%%%%%%%%%%%%%%%%%%%%%%%%%%%%%
%%%%%%%%%%%%%%%%%%%%%%%%%%%%%%%
%%%%%%%%%%%%%%%%%%%%%%%%%%%%%%%
%%%%%%%%%%%%%%%%%%%%%%%%%%%%%%%
%%%%%%%%%%%%%%%%%%%%%%%%%%%%%%%

%\emph{Convex quantum resource theories and resource quantifiers.---}
A general \emph{resource theory} consists of: a set of objects $O$, the identification of a property of these objects to be considered as a resource, and a consequent bipartition of the set of objects into  \emph{resourceful} and \emph{free} objects. If the set of free objects is a convex set, we say that we have a convex resource theory. In this work we focus on the convex QRT of quantum measurements with the resource of informativeness.

\textbf{Definition 1:} (CQRT of measurement informativeness) Consider the set of Positive-Operator Valued Measures (POVMs) acting on a Hilbert space of dimension $d$. A POVM $\mathbb{M}$ is a collection of POVM elements  $\mathbb{M}=\{M_a\}$ with $a\in \{1,...,o\}$ satisfying $M_a\geq 0$ $\forall a$ and $\sum_a M_a=\mathds{1}$.  We now consider the resource of informativeness \cite{RoM}. We say a measurement is uninformative when there exists a probability distribution $q(a)$ such that $M_a=q(a)\mathds{1}$, $\forall a$. We say that the measurement is informative otherwise.

One can check that the set of uninformative measurements forms a convex set and therefore, defines a convex QRT of measurements. It will be useful introduce the notion of simulability of measurements.

\textbf{Definition 2:} (Simulability of measurements \cite{simulability})
We say that a measurement $\mathbb{N}=\{N_x\}$, $x\in \{1,...,k\}$ is simulable by the measurement $\mathbb{M}=\{M_a\}$, $a\in \{1,...,o\}$ when there exists a conditional probability distribution $\{q(x|a)\}$ such that:
\begin{align}
    N_x=\sum_a q(x|a)M_a.
    \label{eq:sim}
\end{align}
One can check that the simulability of measurements defines a partial order for the set of measurements and therefore we use the notation $\mathbb{N} \preceq \mathbb{M}$, meaning that $\mathbb{N}$ is simulable by $\mathbb{M}$. 
Simulability of the measurement $\mathbb{N}$ can be understood as a post-processing of the measurement $\mathbb{M}$. 

We now define a weight-based quantifier for informativeness. The idea is to geometrically quantify the amount of resource contained in an object. This quantifier was originally introduced in \cite{EPR2} in the context of nonlocality and it was later independently rediscovered in \cite{WoE} in the context of entanglement. This quantifier has several different names such as: part, content, cost and weight. In order to keep consistency with recent notation in the literature, we adopt \emph{weight} in this work.

\textbf{Definition 3:} (Weight of informativeness) The weight of informativeness of a measurement $\mathbb{M}=\{M_a\}$ is given by:
\begin{align}
    {\rm WoI}\left(\mathbb{M}\right)= \min_{\substack{w \geq 0\\ \{q(a)\}\\ \mathbb{N}}}
    \left\{ w\,\bigg| \, M_a = wN_a +(1-w)q(a)\mathds{1}\right\}
    \label{eq:WoI}
\end{align}
where $\{q(a)\mathds{1}\}$ is an uninformative measurement and $\mathbb{N}=\{N_a\}$ is a general POVM, $N_a\geq 0, \sum_a N_a=\mathds{1}$. The weight quantifies the minimal amount with which some resourceful measurement $\mathbb{N}$ needs to be used in order to reproduce $\mathbb{M}$. Evaluating the WoI is a semi-definite program (SDP) \cite{Boyd_book} and hence it can be solved efficiently numerically (see Appendix A).

\textbf{Lemma:} (Properties of $\rm WoI$) The weight of informativeness (\autoref{eq:WoI}) satisfies the following properties. (i) \emph{Faithfulness}: 
${\rm WoI}(\mathbb{M})=0 \leftrightarrow \mathbb{M}=\{M_a=q(a)\mathds{1}\}$. (ii) \emph{Convexity}: given two measurements $\mathbb{M}_1, \mathbb{M}_2$ and $p\in[0,1]$ we have ${\rm WoI}\left(p\mathbb{M}_1+(1-p)\mathbb{M}_2\right)\leq p\,{\rm WoI}(\mathbb{M}_1)+(1-p){\rm WoI}(\mathbb{M}_2)$. (iii) \emph{Monotonicity} under measurement simulation: $\mathbb{N} \preceq \mathbb{M} \rightarrow {\rm WoI}(\mathbb{N})\leq {\rm WoI}(\mathbb{N})$. (iv) \emph{Explicit form} ${\rm WoI}(\mathbb{M}) = 1-\sum_a \lambda_\mathrm{min}(M_a)$, where $\lambda_\mathrm{min}(\cdot)$ is the smallest eigenvalue. (v) \emph{Upper bounded by one}: $0\leq {\rm WoI}(\mathbb{M})\leq 1$, $\forall \mathbb{M}$.

The proof of these properties is given in Appendix A and they demonstrate that the WoI is good measure of measurement informativeness. We now show that it also has operational significance, by considering a game known as state exclusion. 

%%%%%%%%%%%%%%%
%%%%%%%%%%%%%%%
%%%%%%%%%%%%%%%
%%%%%%%%%%%%%%%
%%%%%%%%%%%%%%%
%%%%%%%%%%%%%%%
\section{State exclusion games}
%%%%%%%%%%%%%%%
%%%%%%%%%%%%%%%
%%%%%%%%%%%%%%%
%%%%%%%%%%%%%%%
%%%%%%%%%%%%%%%
%%%%%%%%%%%%%%%

%\emph{State exclusion games.---} 
We consider a game first formalised in \cite{CES} for analysing the Pusey-Barrett-Rudolph (PBR) theorem \cite{PBR}. The property considered by PBR has been addressed under different names like antidistinguishability \cite{AD2018} or not-Post-Peierls compatibility (Post-Peierls incompatibility) \cite{DSR2017,PPI2002}. We adopt an operational approach here, so that this property guarantees that the game of state exclusion is won with probability one, or conclusive (or perfect) state exclusion \cite{CES, AM2019}. The game of state exclusion has been explored under noisy channels \cite{AD2018}, as well as its communication complexity properties \cite{CES2,CES3}.

\textbf{Game:} (State exclusion \cite{CES}) A referee has a collection of states $\{\rho_x\}$, $x\in\{1,...,k\}$, and promises to send a player the state $\rho_x$ with probability $p(x)$. The goal is for the player to output a guess $g \in \{1,...,k\}$ of a state that was \textit{not} sent. That is, the player succeeds at the game if $g \neq x$ and fail when $g=x$. A given state exclusion game is fully specified by an ensemble $\mathcal{E} = \{\rho_x,p(x)\}$.

This state exclusion game can be seen as being opposite to the game of state discrimination, in which the goal is to correctly identify the state that was sent. Since the goal is to guess the state that was not sent, this game is referred to as excluding, rather than discriminating. 

We are interested in quantum strategies for the player in this game using a fixed \emph{resourceful} measurement $\mathbb{M}$, and how this compares to the best quantum strategy with \emph{free} measurements (classical strategy). We will quantify how well the player does by the probability of error in excluding a state, which should be as small as possible.

\textbf{Classical and quantum protocols:} The best strategy for a classical player, one that is either unable to perform any quantum measurement, or allowed only to perform uninformative measurements, is easily seen to be to output the index of the least probable state. In this case, the minimal probability of error is:
\begin{align}
    P_{\rm err}^{\rm C}(\mathcal{E})=\min_x \, p(x).
    \label{eq:QSEc}
\end{align}
On the other hand, we consider that the player has the ability to perform a single quantum measurement $\mathbb{M}=\{M_a\}$ with $o$ outcomes. The player could nevertheless simulate a measurement $\mathbb{N}=\{N_x\}$ with $k$ outcomes, according to (\autoref{eq:sim}), and use the measurement result as the guess of which state to exclude. The minimum probability of error following this strategy is then:
\begin{align}
    P_{\rm err}^{\rm Q}(\mathcal{E},\mathbb{M})=
    \min_{\mathbb{N} \preceq \mathbb{M}} \sum_{x} p(x){\rm Tr}[N_x\rho_x],
    \label{eq:QSEq}
\end{align}
with the minimisation being performed over all POVMs $\mathbb{N}$ that are simulable by $\mathbb{M}$ (\autoref{eq:sim}).

We are interested in comparing classical and quantum strategies for different games $\mathcal{E}$. In general the player will have a smaller probability of error using a quantum strategy compared to a classical strategy, and hence $P_{\rm err}^{\rm Q}(\mathcal{E},\mathbb{M}) / p_{\rm err}^\mathrm{C}(\mathcal{E}) \leq 1$. We are interested in the optimal \emph{advantage} that can be  obtained by a fixed measurement $\mathbb{M}$ compared to the best classical strategy, over all games $\mathcal{E}$, i.e. in how small the ratio between quantum and classical error probabilities can be made. In the next section we will show that this is precisely characterised by the weight of informativeness.

%%%%%%%%%%%%%%%%%%%%%%%%%
%%%%%%%%%%%%%%%%%%%%%%%%%
%%%%%%%%%%%%%%%%%%%%%%%%%
%%%%%%%%%%%%%%%%%%%%%%%%%
%%%%%%%%%%%%%%%%%%%%%%%%%
%%%%%%%%%%%%%%%%%%%%%%%%%
\section{Weight of informativeness and state exclusion}
%%%%%%%%%%%%%%%%%%%%%%%%%
%%%%%%%%%%%%%%%%%%%%%%%%%
%%%%%%%%%%%%%%%%%%%%%%%%%
%%%%%%%%%%%%%%%%%%%%%%%%%
%%%%%%%%%%%%%%%%%%%%%%%%%
%%%%%%%%%%%%%%%%%%%%%%%%%

%\emph{Weight of informativeness and state exclusion.---}
In this section we establish a first result relating the weight of informativeness of a measurement with its performance in the game of state exclusion. 

\textbf{Result 1:} Consider a state exclusion game in which the player is sent a state from the ensemble $\mathcal{E}=\{\rho_x,p(x)\}$. The optimal advantage offered by the measurement $\mathbb{M}$ over any classical strategy is given by:
\begin{align}
    \min_{\mathcal{E}} \frac{ P^{\rm Q}_{\rm err}(\mathcal{E},\mathbb{M})}{p_{\rm err}^\mathrm{C}(\mathcal{E})} = 1-{\rm WoI}(\mathbb{M}).
    \label{eq:result1}
\end{align}
This shows that for all exclusion games the WoI bounds the decrease in error probability that can be obtained for any $\mathcal{E}$, and that there exists a game $\mathcal{E}^*$ where this decrease is given precisely by the WoI.

The proof consists of two parts. First we prove that the WoI lower bounds the advantage for all tasks $\mathcal{E}$. Then we prove that this lower bound can be achieved by extracting an optimal ensemble $\mathcal{E}^*$ out of the dual SDP formulation of the WoI. The full proof is given in Appendix B. 

This establishes for the first time an operational interpretation of a weight-based quantifier, making a link to state exclusion, and thus establishing a connection between this two previously unrelated comcepts.

%%%%%%%%%%%%%%%%%%
%%%%%%%%%%%%%%%%%%
%%%%%%%%%%%%%%%%%%
%%%%%%%%%%%%%%%%%%
%%%%%%%%%%%%%%%%%%
%%%%%%%%%%%%%%%%%%
\section{Single-shot information theory}
%%%%%%%%%%%%%%%%%%
%%%%%%%%%%%%%%%%%%
%%%%%%%%%%%%%%%%%%
%%%%%%%%%%%%%%%%%%
%%%%%%%%%%%%%%%%%%
%%%%%%%%%%%%%%%%%%

%\emph{Single-shot information theory.---}
We now analyse the game of state exclusion from a different angle, of a communication task in information theory. Consider a hypothetical situation whereby a person needs to de-activate a bomb, by cutting an appropriate wire. The bomb will only explode if the blue wire is cut -- if any wire is cut it will be deactivated. The person at the bomb doesn't know this, but is on the phone with a knowledgeable person, who tells them what to do. If the phoneline is noisy, what is the safest way to communicate this information? Instead of trying to faithfully communicate `blue' (i.e.~encoding which wire \emph{not} to cut), a better coding strategy may be to communicate as the wire \emph{to} cut, the wire which is least likely to be wrongly decoded as `blue'. 

Thus, in contrast to the usual communication problem, which is about faithfully identifying (or discriminating) information, the above example shows that there are communication problems where the goal is to \emph{exclude} information. The ability of a channel to allow for faithful discrimination may be completely different from its ability to faithfully exclude, and in general different coding strategies should be employed. 

Consider then a random variable $X$, distributed according to $p(x)$, for which an outcome should be successfully excluded, the error probability is $P_\mathrm{err}(X) = \min_x p(x)$. The entropy associated with this error probability is the order minus-infinity R\'enyi entropy, $H_{-\infty}(X) = -\log P_\mathrm{err}(X)$, which we shall call the `exclusion entropy'. Consider a channel specified by the conditional probability distribution $p(y|x)$. The conditional error probability at the outcome of the channel is $P_\mathrm{err}(X|Y) = \sum_y p(y) \min_x p(x|y)$ and the associated conditional exclusion entropy is $H_{-\infty}(X|Y) = -\log P_\mathrm{err}(X|Y)$. The reduction in exclusion entropy is then associated to the mutual exclusion information between $X$ and $Y$, $I_{-\infty}(X:Y) = H_{-\infty}(X|Y)- H_{-\infty}(X)$. 

We can now define the `excludible' information of quantum channel $\Lambda(\cdot)$, but considering optimising over all encodings, i.e. input ensembles $\mathcal{E} = \{p(x),\rho_x \}$, and all decodings, i.e. measurements $\mathbb{D} = \{D_g\}_g$:

\textbf{Definition 4:} The single-shot excludible information of the quantum channel $\Lambda(\cdot)$ is given by:
\begin{align}
    I^{\rm exc}_{-\infty}(\Lambda) &=\max_{\mathcal{E},\mathbb{D}} I_{-\infty}(X:G),
    \label{eq:MIMAI}
\end{align}
where $p(g|x) = \mathrm{Tr}[\Lambda(\rho_x)D_g]$ is the conditional probability distribution of the outcome of the (decoding) measurement, applied to the output of the channel.

We now extend the above weight-exclusion correspondence to a three-way correspondence, by showing that the WoI is also related to the excludible information (\autoref{eq:MIMAI}) of the quantum-to-classical channel $\Lambda_\mathbb{M}(\cdot)$ naturally associated to a measurement via 
\begin{equation}
\Lambda_\mathbb{M}(\rho) = \sum_a \ket{a}\bra{a} \mathrm{Tr}[M_a \rho],
\end{equation}
where $\{\ket{a}\}$ forms an arbitrary basis for the output Hilbert space of the channel.

\textbf{Result 2:} The single-shot excludible information of a quantum-to-classical channel $\Lambda_{\mathbb{M}}$ of the form (\autoref{eq:MIMAI}) is specified by the WoI and is given by:
\begin{align}
    I^{\rm exc}_{-\infty}(\Lambda_{\mathbb{M}})
    =
    -\log \left[1-{\rm WoI}(\mathbb{M})\right].
    \label{eq:result2}
\end{align}
The proof of this result is given in Appendix C. This result parallels the  finding that robustness of informativeness is related to the single-shot accessible (rather than excludible) information of the associated channel,  $I^{\rm acc}_{+\infty}(\Lambda_{\mathbb{M}})=\log\left[1+{\rm RoI}(\mathbb{M})\right]$ (see \cite{RoM} for definitions). 

%%%%%%%%%%%%%%%%%
%%%%%%%%%%%%%%%%%
%%%%%%%%%%%%%%%%%
%%%%%%%%%%%%%%%%%
%%%%%%%%%%%%%%%%%
%%%%%%%%%%%%%%%%%
\section{Complete set of monotones}
%%%%%%%%%%%%%%%%%
%%%%%%%%%%%%%%%%%
%%%%%%%%%%%%%%%%%
%%%%%%%%%%%%%%%%%
%%%%%%%%%%%%%%%%%
%%%%%%%%%%%%%%%%%

%\emph{Complete set of monotones.---}
We have already seen that the simulability of measurements defines a partial order for the set of measurements (\autoref{eq:sim}). We now show that the probabilities of error at the state exclusion game are intimately connected to simulation, providing a complete set of monotones for the partial order. 

\textbf{Result 3:} Consider two measurements $\mathbb{M}$ and $\mathbb{N}$. The measurement $\mathbb{M}$ can simulate the measurement $\mathbb{N}$, $\mathbb{M} \succeq \mathbb{N}$, via (\autoref{eq:sim}), if and only:
\begin{align}
    P_\mathrm{err}^\mathrm{Q}(\mathcal{E},\mathbb{M}) \leq P_\mathrm{err}^\mathrm{Q}(\mathcal{E},\mathbb{N}) \quad \forall \,\mathcal{E} = \{p(x),\rho_x\}.
    \label{eq:result3}
\end{align}
That is, a measurement $\mathbb{M}$ can simulate a measurement $\mathbb{N}$  if and only if it is never worse in any state exclusion game $\mathcal{E}$. The proof of this result is in Appendix D.

This result shows then that the error probabilities over all state exclusion games form a complete set of (decreasing) monotones for the partial order of measurement simulation. It is interesting to note that it was previously shown that the probability of succeeding in state discrimination also forms a complete set of (increasing) monotones for measurement simulation \cite{RoM}. Hence, we now have a second, independent, complete set of monotones. 

%%%%%%%%%%
%%%%%%%%%%
%%%%%%%%%%
%%%%%%%%%%
%%%%%%%%%%
%%%%%%%%%%
\section{Conclusions}
%%%%%%%%%%
%%%%%%%%%%
%%%%%%%%%%
%%%%%%%%%%
%%%%%%%%%%
%%%%%%%%%%

%\emph{Conclusions.---}
In this work we have introduced a weight-based quantifier of measurement informativeness and shown that it has an operational interpretation as the biggest advantage that can be achieved in reducing the error probability in the game of quantum state exclusion. We have furthermore introduced the notions of exclusion-entropy and excludible information associated to a communication task where the information being communicated is naturally related to exclusion rather than identification or discrimination, as is usually the case. We have shown that the weight of informativeness fully characterises the single-shot excludible information of the quantum-to-classical channel associated to a measurement, proving a three-way correspondence, in \emph{parallel} to the one found for the robustness of informativeness \cite{RoM}. Finally, we have shown that exclusion games also constitute a complete set of tasks for measurement simulation, with the error probability over all games forming a complete set of monotones.

Although we have focused here on the quantum resource theory of measurement informativeness, we conjecture that the insight we have found is in fact rather generic for arbitrary quantum resource theories. In particular, we conjecture that whenever a (generalised) robustness-based measure is related to a discrimination task, then a weight-based measure will be related to the corresponding exclusion task, when considering \emph{arbitrary objects} and \emph{arbitrary resources}. In an upcoming paper we provide support to this conjecture by proving that it holds true when considering convex QRTs of states with arbitrary resources \cite{DS2019b}.

%%%%%%%%%%%%%%
%%%%%%%%%%%%%%
%%%%%%%%%%%%%%
%%%%%%%%%%%%%%
%%%%%%%%%%%%%%
%%%%%%%%%%%%%%
\section*{Acknowledgements}
%%%%%%%%%%%%%%
%%%%%%%%%%%%%%
%%%%%%%%%%%%%%
%%%%%%%%%%%%%%
%%%%%%%%%%%%%%
%%%%%%%%%%%%%%

We would like to thank Noah Linden, Patryk Lipka-Bartosik and Tom Purves for insightful discussions. A.F.D acknowledges support from COLCIENCIAS 756-2016. P.S. acknowledges support from a Royal Society URF (UHQT).

\bibliographystyle{apsrev4-1}
\bibliography{bibliography.bib}

\appendix
\section*{APPENDICES}

%%%%%%%%%%%%%%%%%%%%%%%%
%%%%%%%%%%%%%%%%%%%%%%%%
%%%%%%%%%%%%%%%%%%%%%%%%
%%%%%%%%%%%%%%%%%%%%%%%%
%%%%%%%%%%%%%%%%%%%%%%%%
%%%%%%%%%%%%%%%%%%%%%%%%
\section{Lemma}
%%%%%%%%%%%%%%%%%%%%%%%%
%%%%%%%%%%%%%%%%%%%%%%%%
%%%%%%%%%%%%%%%%%%%%%%%%
%%%%%%%%%%%%%%%%%%%%%%%%
%%%%%%%%%%%%%%%%%%%%%%%%
%%%%%%%%%%%%%%%%%%%%%%%%

\textbf{Lemma:} (Properties of $\rm WoI$) The weight of informativeness (\autoref{eq:WoI}) satisfies the following properties:
\begin{enumerate}[(i)]
\item \emph{Faithfulness}: 
\begin{equation}
{\rm WoI}(\mathbb{M})=0 \leftrightarrow \mathbb{M}=\{M_a=q(a)\mathds{1}\}.
\end{equation}
\item \emph{Convexity}: given two measurements $\mathbb{M}_1, \mathbb{M}_2$ and $p\in[0,1]$ we have
\begin{multline}
{\rm WoI}\left(p\mathbb{M}_1+(1-p)\mathbb{M}_2\right)\\\leq p\,{\rm WoI}(\mathbb{M}_1)+(1-p){\rm WoI}(\mathbb{M}_2).
\end{multline}
\item \emph{Monotonicity} (for the order induced by the simulability of measurements):
\begin{equation}
\mathbb{M}' \preceq \mathbb{M}       \quad \rightarrow \quad {\rm WoI}(\mathbb{M}')\leq {\rm WoI}(\mathbb{M}).
\end{equation}
\item \emph{Explicit form}: 
\begin{equation}
{\rm WoI}(\mathbb{M}) = 1-\sum_a \lambda_\mathrm{min}(M_a),
\end{equation}
where $\lambda_\mathrm{min}(\cdot)$ is the smallest eigenvalue.
\item \emph{Upper bounded by one}:
\begin{equation}
0\leq {\rm WoI}(\mathbb{M})\leq 1, \quad \forall \mathbb{M}.
\end{equation}
\end{enumerate} 

\emph{Proof.---}The weight of informativeness of a measurement $\mathbb{M}=\{M_a\}$ is given by
\begin{align*}
    {\rm WoI}\left(\mathbb{M}\right)= \min_{\substack{w \geq 0\\ \{q(a)\}\\ \mathbb{N}}}
    \left\{ w\,\bigg| \, M_a = wN_a +(1-w)q(a)\mathds{1}\right\},
\end{align*}
with $\{q(a)\mathds{1}\}$ an uninformative measurement, $\mathbb{N}=\{N_a\}$ a general POVM, $N_a\geq 0, \sum_a N_a=\mathds{1}$. We address the \emph{optimal triple} associated to ${\rm WoI}(\mathbb{M})=w^*$ as $(w^*,q^*,\mathbb{N}^*)$ so that:
\begin{align}
    M_a=(1-w^*)q^*(a)\mathds{1}+w^*N^*_a, \hspace{0.5cm} \forall a.
    \label{eq:Lemma1}
\end{align} 

%%%%%%%%%%%%%%%%%%
%%%%%%%%%%%%%%%%%% 
\textbf{Part (i).} For the necessary condition we have that if $w^*={\rm WoI}(\mathbb{M})=0$, substituting this in (\autoref{eq:Lemma1}), we have $M_a=q^*(a)\mathds{1}$. For the sufficient condition we have that if $M_a=m(a)\mathds{1}$, we are interested in triples $(w,q,\mathbb{N})$ allowing the decomposition $m(a)\mathds{1}=(1-w)q(a)\mathds{1}+wN_a$. We choose a trial function $q(a):=m(a)$ $\forall a$ for which we have that  $w=0$, which is the minimum possible and so $w^*=w=0$ with $q^*(a)=q(a)$.

%%%%%%%%%%%%%%%%%
%%%%%%%%%%%%%%%%%
\textbf{Part (ii).} Let us consider two measurements $\mathbb{M}_1=\{M_{1a}\}, \mathbb{M}_2=\{M_{2a}\}$ with respective quantities ${\rm WoI}(\mathbb{M}_1)$, ${\rm WoI}(\mathbb{M}_2)$ and their associated optimal triples $(w^*_1,q_1^*,\mathbb{N}_1^*)$ and $(w^*_2,q_2^*,\mathbb{N}_2^*)$ satisfying:
\begin{align*}
    M_{1a}=(1-w_1^*)q_1^*(a)\mathds{1}+w_1^*N^*_{1a},\\
    M_{2a}=(1-w_2^*)q_2^*(a)\mathds{1}+w_2^*N^*_{2a}.
\end{align*} 
We now consider the quantities for $p\in[0,1]$:
\begin{align}
    \nonumber &pM_{1a}+(1-p)M_{2a}=\\
    \nonumber &=p\Bigg[
    (1-w_1^*)q_1^*(a)\mathds{1}+w_1^*N^*_{1a}
    \Bigg]+\\
    &+(1-p)\Bigg[
    (1-w_2^*)q_2^*(a)\mathds{1}+w_2^*N^*_{2a}
    \Bigg].
    \label{eq:rewrite}
\end{align} 
We now define the variables:
\begin{align*}
    \tilde w&=pw_1^*+(1-p)w_2^*,\\
    \tilde q(a)&=\frac{
    p(1-w_1^*)q_1^*(a)+(1-p)(1-w_2^*)q_2^*(a)
    }{
    1-\tilde w
    },\\
    \tilde N_a &=\frac{
    pw_1^*N_{1a}^*+(1-p)w_2^*N_{2a}^*,
    }
    {\tilde w},
\end{align*}
and then we can rewrite (\autoref{eq:rewrite}) as:
\begin{align}
    pM_{1a}+(1-p)M_{2a}=
    (1-\tilde w)\tilde q(a)\mathds{1}+\tilde w \tilde N_a.
    \label{eq:comparing}
\end{align} 
We now consider the quantity ${\rm WoI}\left[p\mathbb{M}^1+(1-p)\mathbb{M}^2\right]$ with associated optimal triple $(W^*,Q^*,\mathbb{N}^*)$ and therefore $\forall a$:
\begin{align}
    pM_{1a}+(1-p)M_{2a}=(1-W^*)Q^*(a)\mathds{1}+W^*N^*_a.
    \label{eq:previous}
\end{align}
Now comparing (\autoref{eq:previous}) with (\autoref{eq:comparing}) we have that:
\begin{align*}
    W^*\leq \tilde w,
\end{align*}
because $W^*$ is the optimal, and therefore obtaining:
\begin{align*}
    {\rm WoI}\left[p\mathbb{M}_1+(1-p)\mathbb{M}_2\right]\leq p\,{\rm WoI}(\mathbb{M}_1)+(1-p){\rm WoI}(\mathbb{M}_2).
\end{align*}

%%%%%%%%%%%%%%%%%%%%
%%%%%%%%%%%%%%%%%%%%
\textbf{Part (iii).} Let us consider that $\mathbb{M}' \preceq \mathbb{M}$ which means:
\begin{align}
    M'_b=\sum_a p(b|a)M_a, \hspace{0.5cm}\forall a.
    \label{eq:simA}
\end{align}
We now consider the quantity ${\rm WoI}(\mathbb{M})$ and its associated optimal triple $(w^*,q^*,\mathbb{N}^*)$ then $\forall a$:
\begin{align}
    M_{a}=(1-w_1^*)q^*(a)\mathds{1}+w_1^*N^*_{a}.
    \label{eq:simB}
\end{align} 
Substituting (\autoref{eq:simB}) in (\autoref{eq:simA}) we have:
\begin{align}
    \nonumber M'_b&=\sum_a p(b|a)M_a,\\
    \nonumber   &=\sum_a p(b|a) \Big[(1-w_1^*)q^*(a)\mathds{1}+w_1^*N^*_{a} \Big],\\
    \nonumber   &=(1-w^*)\sum_a p(b|a)q^*(a)\mathds{1}+w^*\sum_a p(b|a)N^*_{a},\\
    &=(1-w^*)\tilde q(b)\mathds{1}+w^*\tilde N_b, \label{eq:lastline}
\end{align}
where in the last line we have defined the quantities $\tilde q(b)=\sum_ap(b|a)q^*(a)$ and $\tilde N_b=\sum_a p(b|a)N^*_{a}$. We now consider the quantity ${\rm WoI}(\mathbb{M}')$ and its associated optimal triple $(W^*,Q^*, \mathbb{M}^*)$. From (\autoref{eq:lastline}) we have that $w^*$ is a candidate for being $W^*$ but we have that $W^*$ is optimal and therefore $W^*\leq w^*$ which is equivalent to ${\rm WoI}(\mathbb{M}')\leq {\rm WoI}(\mathbb{M})$.

%%%%%%%%%%%%%%%%%%%
%%%%%%%%%%%%%%%%%%%
\textbf{Part (iv) and (v).} By definition we have that ${\rm WoI}(\mathbb{M})\geq 0$ so we now check the upper bound. Let us start again with the weight of informativeness of a measurement $\mathbb{M}=\{M_a\}$. Renaming $\tilde N_a=wN_a$ and $\tilde q(a)=(1-w)q(a)$ we have that $\forall a$:
\begin{align}
     M_a-\tilde q(a)\mathds{1}&=\tilde N_a\geq 0.
     \label{eq:I0}
\end{align}
Minimising $w$ is equivalent to maximising $(1-w)$ and together with $\sum_a \tilde q(a)=1-w$ we have: 
\begin{align*}
    1-{\rm WoI}(\mathbb{M})=\max_{w\geq 0}\, \left\{1-w\right\}=\max_{\tilde q}\sum^{o}_{a=1} \tilde q(a).
\end{align*}
We can now explicitly define a primal SDP as:
\begin{align}
   \nonumber 1-{\rm WoI}(\mathbb{M})=
    \max_{\tilde q} &\sum^{o}_{a=1} \tilde q(a),\\
    \text{s.t. } & M_a-\tilde q(a) \mathds{1}\geq 0, \hspace{0.3cm} \forall a. \label{eq:Primal}
\end{align}
With the later inequality being the constraint (\autoref{eq:I0}). The constraint means that $M_a \geq \tilde q(a) \mathds{1}$ and so ${\rm max}_{\tilde q} \sum^{o}_{a=1} \tilde q(a)=\sum_a \lambda_{\rm min}(M_a)$ with $\lambda_{\rm min}(M_a)$ the smallest eigenvalue of $M_a$ and therefore:
\begin{align*}
   \nonumber {\rm WoI}(\mathbb{M})=1-\sum_a \lambda_{\rm min}(M_a).
\end{align*}
The operators $M_a$ are POVM elements, $M_a\geq 0$, which means that $\lambda_{\rm min}(M_a)\geq 0$ and so ${\rm WoI}(\mathbb{M})\leq 1$.  The upper bound is achieved by any measurement such that all the POVM elements are non-full-rank. For example, a rank-1 (projective) measurement $\Pi=\{\Pi_a\}$, $\Pi_a \geq 0$, $\sum_a \Pi_a=\mathds{1}$, $\Pi_a \Pi_b=\delta_{ab}\Pi_{a}$ has maximal weight of informativeness, since $\lambda_{\rm min}(\Pi_a)=0$ $\forall a$ and therefore ${\rm WoI}(\Pi)=1$.

%%%%%%%%%%%%%%%%%%%%%%%%%%%
%%%%%%%%%%%%%%%%%%%%%%%%%%%
%%%%%%%%%%%%%%%%%%%%%%%%%%%
%%%%%%%%%%%%%%%%%%%%%%%%%%%
%%%%%%%%%%%%%%%%%%%%%%%%%%%
%%%%%%%%%%%%%%%%%%%%%%%%%%%
\section{Proof of Result 1}
%%%%%%%%%%%%%%%%%%%%%%%%%%%
%%%%%%%%%%%%%%%%%%%%%%%%%%%
%%%%%%%%%%%%%%%%%%%%%%%%%%%
%%%%%%%%%%%%%%%%%%%%%%%%%%%
%%%%%%%%%%%%%%%%%%%%%%%%%%%
%%%%%%%%%%%%%%%%%%%%%%%%%%%

In the appendix we prove Result 1 of the main text. We prove the result in two parts. We first prove the lower bound, and then we prove that it can be achieved.

%%%%%%%%%%%%%%%%%%%%%%%
%%%%%%%%%%%%%%%%%%%%%%%
%%%%%%%%%%%%%%%%%%%%%%%
%%%%%%%%%%%%%%%%%%%%%%%
\subsection{First part}
%%%%%%%%%%%%%%%%%%%%%%%
%%%%%%%%%%%%%%%%%%%%%%%
%%%%%%%%%%%%%%%%%%%%%%%
%%%%%%%%%%%%%%%%%%%%%%%

In this first part we prove that:
\begin{align}
    [1-{\rm WoI}(\mathbb{M})]
    P_{\rm err}^{\rm C}(\mathcal{E}) 
    \leq
    P_{\rm err}^{\rm Q}(\mathcal{E},\mathbb{M}), \hspace{0.3cm} \forall \mathcal{E}, \mathbb{M}.
    \label{eq:I1}
\end{align}
Let us start with the weight of informativeness of a measurement as  given by (\autoref{eq:WoI}). Consider that the minimum is achieved with the triple $(q^*, \mathbb{N}^*, w^*)$ so that $\forall a$:
\begin{align}
    \nonumber M_a-(1-w^*)q^*(a)\mathds{1}=w^*N_a\geq 0, 
\end{align} 
which implies that
\begin{equation}\label{eq:ineq}
M_a\geq \left[1-{\rm WoI}\left(\mathbb{M}\right)\right]q^*(a)\mathds{1}.
\end{equation}
where we use the fact that $w^*={\rm WoI}\left(\mathbb{M}\right)$. We now address the probability of error in state exclusion:
\begin{align*}
    &P_{\rm err}^{\rm Q}(\mathcal{E},\mathbb{M})\\ &=\min_{\mathbb{M} \succeq \mathbb{N}} \sum^k_{x=1} {\rm Tr}\left(N_x\tilde\rho_x\right),\\
    &=\min_{\{p(x|a)\}} \sum^k_{x=1} {\rm Tr}\left\{\left[  \sum^{o}_{a=1} p(x|a) M_a \right]\tilde \rho_x \right \},\\
    &\geq \min_{\{p(x|a)\}} \sum^{k}_{x=1} {\rm Tr}\left\{\left[  \sum^{o}_{a=1} p(x|a) \left[(1-{\rm WoI}(\mathbb{M}))q(a)\mathds{1}\right] \right]\tilde \rho_x \right \},\\
    &= \min_{\{p(x|a)\}} \sum^k_{x=1} \sum^{o}_{a=1}
    p(x)p(x|a) (1-{\rm WoI}(\mathbb{M}))q(a),\\
    &= (1-{\rm WoI}(\mathbb{M})) \min_{\{p(x|a)\}}  \sum^k_{x=1} \sum^{o}_{a=1} p(x)  p(x|a) q(a).
\end{align*}
We use $\tilde \rho_x=p(x)\rho_x$. In the third line we used the inequality (\autoref{eq:ineq}). We now use the fact that $p(x)\geq P_{\rm err}^{\rm C}(\mathcal{E})$, $\forall x$ and that $\sum_x p(x|a)=1, \forall a$ and so we obtain:
\begin{align*}
    &P_{\rm err}^{\rm Q}(\mathcal{E},\mathbb{M})\\
    &\geq (1-{\rm WoI}(\mathbb{M})) \min_{\{p(x|a)\}}  \sum^k_{x=1} \sum^{o}_{a=1} 
    P_{\rm err}^{\rm C}(\mathcal{E})  
    p(x|a) q(a),\\
    &= (1-{\rm WoI}(\mathbb{M}))
    P_{\rm err}^{\rm C}(\mathcal{E})  
    \sum^{o}_{a=1} q(a), \\
    &= (1-{\rm WoI}(\mathbb{M}))
    P_{\rm err}^{\rm C}(\mathcal{E}). 
\end{align*}

%%%%%%%%%%%%%%%%%%%%%%%%
%%%%%%%%%%%%%%%%%%%%%%%%
%%%%%%%%%%%%%%%%%%%%%%%%
%%%%%%%%%%%%%%%%%%%%%%%%
\subsection{Second part}
%%%%%%%%%%%%%%%%%%%%%%%%
%%%%%%%%%%%%%%%%%%%%%%%%
%%%%%%%%%%%%%%%%%%%%%%%%
%%%%%%%%%%%%%%%%%%%%%%%%

In this second part we prove that $\forall \mathbb{M}$, $\exists \mathcal{E}^{\mathbb{M}}$ such that:
\begin{align}
    [1-{\rm WoI}(\mathbb{M})]
    P_{\rm err}^{\rm C}\left(\mathcal{E}^{\mathbb{M}}\right)
    \geq
    P_{\rm err}^{\rm Q}(\mathcal{E}^{\mathbb{M}},\mathbb{M}), \quad \forall \mathbb{M}.
    \label{eq:I2}
\end{align}
This will be done by considering the dual formulation of the primal SDP for the weight of informativeness \cite{Boyd_book}. 

%%%%%%%%%%%%%%%%%%%%%%%%%%%%%%%%%%%%%
%%%%%%%%%%%%%%%%%%%%%%%%%%%%%%%%%%%%%
\subsubsection{Deriving the dual SDP}
%%%%%%%%%%%%%%%%%%%%%%%%%%%%%%%%%%%%%
%%%%%%%%%%%%%%%%%%%%%%%%%%%%%%%%%%%%%

We start by addressing the primal sdp for the weight of informativeness (\autoref{eq:Primal}). We want to maximise the function $f=\sum^{o}_{a=1} \tilde q(a)$ under the constraints that $M_a-\tilde q(a) \mathds{1}\geq 0$ $\forall a$ which is equivalent to the constraint that $\forall \{\rho_a \geq 0\}$  ${\rm Tr} \left [\rho_a (M_a-\tilde q(a) \mathds{1}) \right]\geq 0$ which implies that $\sum_a {\rm Tr} \left [\rho_a (M_a-\tilde q(a) \mathds{1}) \right]\geq 0$. We now write the Lagrangian function using this last constraint as:
\begin{align}
    L= \sum^{o}_{a=1} \tilde q(a)+
    \sum^{o}_{a=1} {\rm Tr}\{\rho_a[M_a-\tilde q(a)\mathds{1}]\}.
    \label{eq:L}
\end{align} 
Let us first note that by construction we have that:
\begin{align}
    L \geq \sum^{o}_{a=1} \tilde q(a).
    \label{eq:bydef}
\end{align}
We now rearrange (\autoref{eq:L}) to get:
\begin{align*}
    L= \sum^{o}_{a=1} \tilde q(a)[1-{\rm Tr}(\rho_a)]+\sum^{o}_{a=1} {\rm Tr}(\rho_aM_a).
\end{align*}
Imposing the condition $1-{\rm Tr}(\rho_a)=0$ $\forall a$ we have that:
\begin{align*}
    L = \sum^{o}_{a=1} {\rm Tr}(\rho_a M_a).
\end{align*}
Using this together with (\autoref{eq:bydef}) we have:
\begin{align*}
    L = \sum^{o}_{a=1} {\rm Tr}(\rho_a M_a) \geq \sum^{o}_{a=1} \tilde q(a).
\end{align*}
Considering now maximising over $\{ \tilde q \}$ we see that
\begin{align*}
    L = \sum^{o}_{a=1} {\rm Tr}(\rho_a M_a) \geq \max_{\tilde q}\sum^{o}_{a=1} \tilde q(a)=1-{\rm WoI}(\mathbb{M}).
\end{align*}
Furthermore, by minimising  over $\{\rho_a\}$, and by \emph{strong duality} \cite{Boyd_book}, which guarantees the equality, we have:
\begin{align*}
    \min_{\{\rho_a\}} L
    &=\min_{\{\rho_a\}}\sum^{o}_{a=1} {\rm Tr}(\rho_a M_a), \\
    &= \max_{\tilde q}\sum^{o}_{a=1} \tilde q(a)=1-{\rm WoI}(\mathbb{M}).
\end{align*}
We then have the dual SDP of (\autoref{eq:Primal}):
\begin{align}
    \nonumber 1-{\rm WoI}(\mathbb{M})=
    &\min_{\{\rho_a\}} \sum^{o}_{a=1} {\rm Tr}(\rho_a M_a),\\
    &\text{s.t.} \rho_a \geq 0, \quad {\rm Tr}(\rho_a)=1\hspace{0.3cm} \forall a.
    \label{eq:Dual}
\end{align}
This dual SDP is going to be useful in what follows.

%%%%%%%%%%%%%%%%%%%%%%%%%%%%%%%%%%%%%
%%%%%%%%%%%%%%%%%%%%%%%%%%%%%%%%%%%%%
\subsubsection{Achieving lower bound}
%%%%%%%%%%%%%%%%%%%%%%%%%%%%%%%%%%%%%
%%%%%%%%%%%%%%%%%%%%%%%%%%%%%%%%%%%%%

We now claim that the optimal ensemble (for achieving the lower bound in (\autoref{eq:I2}) is given by $\mathcal{E}^{\mathbb{M}}=\left\{\rho^{\mathbb{M}}_a,\frac{1}{o} \right\}$, $a=1,...,o$, $P^{\rm C}_{\rm err}\left(\mathcal{E}^{\mathbb{M}}\right)=\frac{1}{o}$ and $\{\rho^{\mathbb{M}}_a\}$ the set of operators coming from the dual SDP (\autoref{eq:Dual}) for a given $\mathbb{M}$. The set $\left\{\rho_a^{\mathbb{M}}\right\}$ then satisfies:
\begin{align*}
    1-{\rm WoI}(\mathbb{M})=\sum^{o}_{a=1} {\rm Tr}\left(
    \rho^{\mathbb{M}}_a M_a
    \right).
\end{align*}
The probability of error in quantum state exclusion for the ensemble $\mathcal{E}^{\mathbb{M}}$ and the measurement $\mathbb{M}$ is then given by:
\begin{align*}
    P_{\rm err}^{\rm Q}(\mathcal{E}^{\mathbb{M}},\mathbb{M})&=
    \min_{\mathbb{N}\prec\mathbb{M}} \sum^{o}_{a=1} {\rm Tr} \left(
    N_a \rho_a^{\mathbb{M}} \frac{1}{o}
    \right),\\
    &=
    \, \min_{\mathbb{N}\prec\mathbb{M}}  \frac{1}{o} \sum^{o}_{a=1} {\rm Tr} \left(
     N_a \rho_a^{\mathbb{M}}
     \right).
\end{align*}  
Given the measurement $\mathbb{M}$, we now choose \emph{not} to simulate any measurement $\mathbb{N}$ but to play with $\mathbb{M}$ instead so:
\begin{align*}
    \hspace{2.5cm}&\leq \frac{1}{o}\sum^{o}_{a=1} {\rm Tr} \left(
    M_a \rho_a^{\mathbb{M}}
    \right),\\
    &=\frac{1}{o}[1-{\rm WoI}(\mathbb{M})],\\
    &=
    P^{\rm C}_{\rm err}\left(\mathcal{E}^{\mathbb{M}}\right)
    [1-{\rm WoI}(\mathbb{M})].
\end{align*}
Putting together the inequalities (\autoref{eq:I1}) and (\autoref{eq:I2}) we obtain the claim in Result 1:
\begin{align*}
    1-{\rm WoI}(\mathbb{M})= \min_{\mathcal{E}} \frac{ P^{\rm Q}_{\rm err}(\mathcal{E},\mathbb{M})}{
    P_{\rm err}^{\rm C}(\mathcal{E})
    }.
\end{align*} 

%%%%%%%%%%%%%%%%%%%%%%%%%%%
%%%%%%%%%%%%%%%%%%%%%%%%%%%
%%%%%%%%%%%%%%%%%%%%%%%%%%%
%%%%%%%%%%%%%%%%%%%%%%%%%%%
%%%%%%%%%%%%%%%%%%%%%%%%%%%
%%%%%%%%%%%%%%%%%%%%%%%%%%%
\section{Proof of Result 2}
%%%%%%%%%%%%%%%%%%%%%%%%%%%
%%%%%%%%%%%%%%%%%%%%%%%%%%%
%%%%%%%%%%%%%%%%%%%%%%%%%%%
%%%%%%%%%%%%%%%%%%%%%%%%%%%
%%%%%%%%%%%%%%%%%%%%%%%%%%%
%%%%%%%%%%%%%%%%%%%%%%%%%%%

In this appendix we calculate the the single-shot excludible information, which we show is specified in terms of the weight of informativeness. In particular, 
\begin{align}
    I^{\rm exc}_{-\infty}(\Lambda_{\mathbb{M}}) =\max_{\mathcal{E},\mathbb{D}}\, I_{-\infty}(X:G),
    \label{eq:B0}
\end{align}
with the mutual exclusion information:
\begin{align}
    I_{-\infty}(X:G)=H_{-\infty}(X|G)-H_{-\infty}(X),
    \label{eq:B1}
\end{align}
and the exclusion entropy and conditional entropy given by:
\begin{align}
    H_{-\infty}(X)&=-\log \min_x p(x) =-\log P_{\rm err}^{\rm C}(\mathcal{E}), \label{eq:B2} \\
    H_{-\infty}(X|G)&=-\log \sum_g \min_x p(x,g),
\end{align}
with $p(x,g)=p(x)p(g|x)$ and $p(g|x)={\rm Tr}\left[\Lambda_{\mathbb{M}}(\rho_x)D_g\right]=\sum_a {\rm Tr}(M_a\rho_x) \braket{a|D_g|a}$. Choosing $D_g=\ketbra{g}{g}$ so that $\braket{a|D_g|a}=\delta^a_x$ and substituting we have:
\begin{align}
    H_{-\infty}(X|G)
    =- \log 
    \sum_g \min_x  
    p(x) \sum_a {\rm Tr}(M_a\rho_x) \delta^a_g \nonumber \\
    =- \log 
        \sum_g \min_x  
        p(x) {\rm Tr}(M_g\rho_x)  .
\end{align} 
Considering $f_g(x)=p(x){\rm Tr}(M_g \rho_x)$ and using:
\begin{align*}
    \min_x f_g(x) =\min_{\{p(x|g)\}} \sum_x p(x|g) f_g(x),
\end{align*}
we have:
\begin{align*} 
    H_{-\infty}(X|G)&=- \log 
    \sum_g \min_{\{p(x|g)\}} \sum_x p(x|g) 
    f_g(x)
    ,\\
    &=- \log 
    \sum_g \min_{\{p(x|g)\}} \sum_x p(x|g) p(x){\rm Tr}(M_g \rho_x).
\end{align*}
Denoting $\tilde \rho_x= p(x)\rho_x$, and re-arranging, this is equivalent to
\begin{align}
    H_{-\infty}(X|G)&=- \log 
    \nonumber \min_{\{p(x|g)\}} \sum_x p(x|g) \sum_g {\rm Tr}(M_g\tilde \rho_x),\\
    \nonumber &=-\log \min_{\{p(x|g)\}} \sum_x {\rm Tr}\left[\left( \sum_g p(x|g) M_g \right) \tilde \rho_x \right],\\
    \nonumber &= -\log  
    \min_{\mathbb{N}\prec\mathbb{M}} \sum_x {\rm Tr}(N_x \tilde \rho_x) 
    ,\\
    &= -\log P^{\rm Q}_{\rm err}(\mathcal{E},\mathbb{M}).
    \label{eq:B3}
\end{align}
Combining (\autoref{eq:B3}) and (\autoref{eq:B2}) with (\autoref{eq:B1}) we obtain:
\begin{align}
    I_{-\infty}(X:G)= \log \left[ \frac{
    P_{\rm err}^{\rm C}(\mathcal{E})
    }{
    P^{\rm Q}_{\rm err}(\mathcal{E},\mathbb{M})
    }
    \right].
    \label{eq:B4}
\end{align}
Substituting now (\autoref{eq:B4}) into (\autoref{eq:B0}) we have:
\begin{align*}
    I^{\rm exc}_{-\infty}(\Lambda_{\mathbb{M}}) 
    &=\max_{\mathcal{E}, \mathbb{D}}\, I_{-\infty}(X:G),\\
    &=\max_{\mathcal{E}, \mathbb{D}} \,
    \log \left \{ 
    \frac{
    P_{\rm err}^{\rm C}(\mathcal{E})
    }{
    P^{\rm Q}_{\rm err}(\mathcal{E},\mathbb{M})
    } 
    \right\},\\
    &=\max_{\mathcal{E}, \mathbb{D}} \,
    -\log \left \{ 
    \frac{
    P^{\rm Q}_{\rm err}(\mathcal{E},\mathbb{M})
    }{
    P_{\rm err}^{\rm C}(\mathcal{E})
    } 
    \right\},\\
    &=-\min_{\mathcal{E}, \mathbb{D}} \,
    \log \left \{ 
    \frac{
    P^{\rm Q}_{\rm err}(\mathcal{E},\mathbb{M})
    }{
    P_{\rm err}^{\rm C}(\mathcal{E})
    } 
    \right\},\\
    &= -\log \left\{ 
    \min_{\mathcal{E}, \mathbb{D}} \left[ \frac{
    P^{\rm Q}_{\rm err}(\mathcal{E},\mathbb{M})
    }{
    P_{\rm err}^{\rm C}(\mathcal{E})
    }\right] 
    \right\},\\
    &=-\log\left[1-{\rm WoI}(\mathbb{M})\right].
\end{align*}
In the last line we have used Result 1 (\autoref{eq:result1}).

%%%%%%%%%%%%%%%%%%%%%%%%%%%
%%%%%%%%%%%%%%%%%%%%%%%%%%%
%%%%%%%%%%%%%%%%%%%%%%%%%%%
%%%%%%%%%%%%%%%%%%%%%%%%%%%
%%%%%%%%%%%%%%%%%%%%%%%%%%%
%%%%%%%%%%%%%%%%%%%%%%%%%%%
\section{Proof of Result 3}
%%%%%%%%%%%%%%%%%%%%%%%%%%%
%%%%%%%%%%%%%%%%%%%%%%%%%%%
%%%%%%%%%%%%%%%%%%%%%%%%%%%
%%%%%%%%%%%%%%%%%%%%%%%%%%%
%%%%%%%%%%%%%%%%%%%%%%%%%%%
%%%%%%%%%%%%%%%%%%%%%%%%%%%

In this appendix we prove that the error probability forms a complete set of monotones for measurement simulation. We do this by showing both necessary and sufficient conditions. 

%%%%%%%%%%%%%%%%%%%%%%%%%%%%%%%%
%%%%%%%%%%%%%%%%%%%%%%%%%%%%%%%%
%%%%%%%%%%%%%%%%%%%%%%%%%%%%%%%%
%%%%%%%%%%%%%%%%%%%%%%%%%%%%%%%%
\subsection{Necessary condition}
%%%%%%%%%%%%%%%%%%%%%%%%%%%%%%%%
%%%%%%%%%%%%%%%%%%%%%%%%%%%%%%%%
%%%%%%%%%%%%%%%%%%%%%%%%%%%%%%%%
%%%%%%%%%%%%%%%%%%%%%%%%%%%%%%%%

Let us address the necessary condition:
\begin{align}
    \mathbb{M} \succeq \mathbb{M}' \implies P_\mathrm{err}^\mathrm{Q}(\mathcal{E},\mathbb{M}) \leq P_\mathrm{err}^\mathrm{Q}(\mathcal{E},\mathbb{M}') \quad  \forall \mathcal{E}.
    \label{eq:C1}
\end{align}
Let us consider the probability of error in state exclusion:
\begin{align*}
    P_{\rm err}^{\rm Q}(\mathcal{E},\mathbb{M'})
    &=\min_{\mathbb{M}' \succeq \mathbb{N}'} \sum^k_{x=1} {\rm Tr}\left(N'_x\tilde\rho_x\right),\\
    &=\min_{\{p(x|b)\}} \sum^k_{x=1} {\rm Tr}\sum^{l'}_{b=1} p(x|b) M'_b \tilde \rho_x,\\
    &=\min_{\{p(x|b)\}} \sum^k_{x=1} {\rm Tr}\sum^{l'}_{b=1} p(x|b) \sum^{l}_{a=1} q(b|a)M_a \tilde \rho_x,\\
    &=\min_{\{p(x|b)\}}\sum^k_{x=1} {\rm Tr}\left[ \sum^{l}_{a=1} r(x|a) M_a \right]\tilde \rho_x.
\end{align*}
In the third line we have used the fact that $\mathbb{M} \succeq \mathbb{M}'$ which means that $M'_b=\sum^{l}_{a=1} q(b|a)M_a$, $\forall b$. We furthermore introduced the conditional probability $\{r(x|a)\}$ such that:
\begin{align*}
  r(x|a)=\sum^{l'}_{b=1} p(x|b)  q(b|a).
\end{align*}
This may not be the most general set of conditional probabilities, therefore
\begin{align*}
    P_{\rm err}^{\rm Q}(\mathcal{E},\mathbb{M'})
    &\geq \min_{\{p(x|a)\}} \sum^k_{x=1} {\rm Tr}\sum^l_{a=1} p(x|a) M_a \tilde \rho_x,\\
    &=\min_{\mathbb{M} \succeq \mathbb{N}} \sum^k_{x=1} {\rm Tr}\left(N_x\tilde\rho_x\right),\\
    &=P_{\rm err}^{\rm Q}(\mathcal{E},\mathbb{M}),
\end{align*}
and therefore obtaining:
\begin{align*}
    P_{\rm err}^{\rm Q}(\mathcal{E},\mathbb{M}')
    \geq 
    P_{\rm err}^{\rm Q}(\mathcal{E},\mathbb{M}),
\end{align*}
as required. 

%%%%%%%%%%%%%%%%%%%%%%%%%%%%%%%%%
%%%%%%%%%%%%%%%%%%%%%%%%%%%%%%%%%
%%%%%%%%%%%%%%%%%%%%%%%%%%%%%%%%%
%%%%%%%%%%%%%%%%%%%%%%%%%%%%%%%%%
\subsection{Sufficient condition}
%%%%%%%%%%%%%%%%%%%%%%%%%%%%%%%%%
%%%%%%%%%%%%%%%%%%%%%%%%%%%%%%%%%
%%%%%%%%%%%%%%%%%%%%%%%%%%%%%%%%%
%%%%%%%%%%%%%%%%%%%%%%%%%%%%%%%%%

We now address the sufficient condition:
\begin{align}
    \mathbb{M} \succeq \mathbb{M}' \impliedby P_\mathrm{err}^\mathrm{Q}(\mathcal{E},\mathbb{M}) \leq 
    P_\mathrm{err}^\mathrm{Q}(\mathcal{E},\mathbb{M}') \quad  \forall \mathcal{E}.
    \label{eq:C1}
\end{align}
Let us start by assuming that the right-hand side is true. 
We now want to prove that $\mathbb{M} \succeq \mathbb{M}'$ which is equivalent to $\sum_a q(x|a)M_a = M_x'$. Let us continue by considering the inequality:
\begin{align}
    \nonumber 0& \geq P_{\rm err}^{\rm Q}(\mathcal{E},\mathbb{M})
    -P_{\rm err}^{\rm Q}(\mathcal{E},\mathbb{M}'), \quad \forall \mathcal{E}    \\
    \nonumber &=\min_{\mathbb{N} \preceq \mathbb{M}} \sum^k_{x=1} {\rm Tr}\left(N_x\tilde\rho_x\right)-    \min_{\mathbb{N}' \preceq \mathbb{M}'} \sum^k_{x=1} {\rm Tr}\left(N'_x\tilde\rho_x\right),\\
%%%%%%%%%%%%%%%%%%%%%%%%%%%%%%%%%%
   \nonumber &\geq \min_{\mathbb{N} \preceq \mathbb{M}} \sum^k_{x=1} {\rm Tr}\left(N_x\tilde\rho_x\right)- \sum^k_{x=1} {\rm Tr}\left(M'_x\tilde\rho_x\right),\\
    \nonumber &= \min_{\mathbb{N} \preceq \mathbb{M}} \sum^k_{x=1} {\rm Tr}\left[\left(N_x-M'_x\right)\tilde\rho_x\right],\\
    &= \min_{\mathbb{N} \preceq \mathbb{M}} \sum^k_{x=1} {\rm Tr}\left[\left( \sum_a p(x|a)M_a-M'_x \right)\tilde\rho_x\right].  \label{eq:quantity}
\end{align}
In the third line we have chosen \emph{not} to simulate any measurement $\mathbb{N}'$ but to keep $\mathbb{M}'$. Let us now define the operators and the magnitude: 
\begin{align}
    \Delta_x (\mathbb{M},\mathbb{M}')
    &=\sum_a p(x|a)M_a-M'_x, \hspace{0.3cm} \forall x,
    \label{eq:Delta_x}\\
    \Delta (\mathcal{E},\mathbb{M},\mathbb{M}')
    &=\sum^k_{x=1} {\rm Tr}\Big[ 
    \Delta_x (\mathbb{M},\mathbb{M}') \tilde\rho_x
    \Big].
    \label{eq:Delta}
\end{align}
Then the quantity in (\autoref{eq:quantity}) becomes:
\begin{align}
    \nonumber &0 \geq \min_{\mathbb{N} \preceq \mathbb{M}} \,
    \Delta (\mathcal{E},\mathbb{M},\mathbb{M}').
\end{align}
This last equation is valid $\forall \mathcal{E}$ and therefore it is in particular, valid for the ensemble that maximises the magnitude:
\begin{align}
    \nonumber 
    &0 \geq 
    \max_{\mathcal{E}} \,
    \min_{\mathbb{N} \preceq \mathbb{M}}\, 
    \Delta (\mathcal{E},\mathbb{M},\mathbb{M}')
    ,\\
    &0 \geq 
    \min_{\mathbb{N} \preceq \mathbb{M}} \,
    \max_{\mathcal{E}} \,
    \Delta (\mathcal{E},\mathbb{M},\mathbb{M}'),
    \label{eq:contra}
\end{align}
where we have used the minimax theorem. If $\Delta_x=\hat 0$ $\forall x$, we have already obtain the desired result that $\sum_a p(x|a)M_a=M'_x$. The idea now is to prove that if we assume otherwise, we obtain a contradiction. We then assume that:
\begin{align}
    \Delta_x(\mathbb{M},\mathbb{M}')=\left(\sum_a p(x|a)M_a-M'_x \right) \neq \hat0, \hspace{0.2cm} \forall x.
    \label{eq:assumption}
\end{align}
One can directly check that we also have:
\begin{align}
    \sum^k_{x=1} \Delta_x(\mathbb{M},\mathbb{M}')=\hat0.  
    \label{eq:condition}
\end{align}
It follows then that i) the operators $\{\Delta_x\}$ cannot \emph{all} be positive, since this would be in contradiction to (\autoref{eq:condition}) ii) $\{\Delta_x\}$ cannot \emph{all} be negative, since this also leads to a contradiction with (\autoref{eq:condition}) iii) $\{\Delta_x\}$ cannot \emph{all} be the zero operator (by assumption (\autoref{eq:assumption})). Therefore, the set $\{\Delta_x\}$ has to contain \emph{at least}: one positive and one negative operator. Let us consider the positive operator. There exists then at least one $x$, say $x^*$, such that $\Delta_{x^*} > \hat 0$, which means that it has to have at least one positive eigenvalue $\lambda_{x^*}^{\rm pos}>0$ with eigenvector $\ket{\lambda_{x^*}^{\rm pos}}$. $\Delta_{x^*}$ is a Hermitian operator and therefore is diagonalisable as $\Delta_{x^*}=\sum_i \lambda^i \ketbra{\lambda_{x^*}^i}{\lambda_{x^*}^i}$ with $\{\ket{\lambda_{x^*}^i}\}$ forming an orthonormal basis. Equivalently, we can write this as:
\begin{align}
    \Delta_{x^*}=\lambda_{x^*}^{\rm pos} \ketbra{\lambda_{x^*}^{\rm pos}}{\lambda_{x^*}^{\rm pos}}+
    \sum_{i\neq {\rm pos}} \lambda_{x^*}^i \ketbra{\lambda_{x^*}^i}{\lambda_{x^*}^i}.
    \label{eq:Delta_x*}
\end{align}
We now consider an ensemble $\mathcal{E}^*=\{\delta^{x^*}_x,\rho_x\}$ with $\rho_{x^*}=\ketbra{\lambda_{x^*}^{\rm pos}}{\lambda_{x^*}^{\rm pos}}$, and the rest of states being arbitrary. With this ensemble we calculate the quantity in (\autoref{eq:Delta})
\begin{align*}
    \Delta (\mathcal{E}^*,\mathbb{M},\mathbb{M}')
    &= \sum^k_{x=1} {\rm Tr}\left[
    \Delta_{x^*}
    \ketbra{\lambda_{x^*}^{\rm pos}}{\lambda_{x^*}^{\rm pos}}
    \delta^{x^*}_x
    \right ],\\
    &= {\rm Tr}\Big[
    \Delta_{x^*}
    \ketbra{\lambda_{x^*}^{\rm pos}}{\lambda_{x^*}^{\rm pos}}
    \Big],\\
    &=\lambda_{x^*}^{\rm pos}>0.
\end{align*}
This is in contradiction with (\autoref{eq:contra}). This follows because from \eqref{eq:contra} we have $\Delta(\mathcal{E}^*)\leq \max_{\mathcal{E}}\Delta(\mathcal{E})\leq 0$. Therefore, the assumption made in \eqref{eq:assumption} is not true, which means that:
\begin{align*}
    &\Delta_x(\mathbb{M},\mathbb{M}')=\sum_a p(x|a)M_a-M'_x = \hat0, \hspace{0.3cm} \forall x,
\end{align*}
from which we obtain
\begin{align*}
    &M'_x=\sum_a p(x|a)M_a,
\end{align*}
or that $\mathbb{M}$ simulates $\mathbb{M}'$, $\mathbb{M}\succeq \mathbb{M}'$.

\end{document}